\documentclass[12pt,preprint]{aastex}

\usepackage{graphicx}
\usepackage{natbib}
\usepackage{amsmath}

\shorttitle{NDAF model for X-ray flares}
\shortauthors{Luo et al.}

\begin{document}

\title{Neutrino-cooled Accretion Model with Magnetic Coupling \\
for X-ray Flares in GRBs}

\author{Yang Luo, Wei-Min Gu, Tong Liu, and Ju-Fu Lu}

\affil{Department of Astronomy and Institute of Theoretical Physics
and Astrophysics, \\
Xiamen University, Xiamen, Fujian 361005, China}

\begin{abstract}
The neutrino-cooled accretion disk, which was proposed to work as
the central engine of gamma-ray bursts, encounters difficulty in
interpreting the X-ray flares after the prompt gamma-ray emission.
In this paper, the magnetic coupling between the inner disk
and the central black hole is taken into consideration.
For mass accretion rates around $0.001 \sim 0.1$~$M_{\sun}$~s$^{-1}$,
our results show that the luminosity of neutrino annihilation
can be significantly enhanced due to the coupling effects.
As a consequence, after the gamma-ray emission, a remnant disk with
mass $M_{\rm disk} \la 0.5$~$M_{\sun}$ may power most of the observed
X-ray flares with the rest frame duration less than 100 seconds.
In addition, a comparison between the magnetic coupling process
and the Blandford-Znajek mechanism is shown on the extraction
of black hole rotational energy.
\end{abstract}   


\section{Introduction}

The launch of Swift satellite has led to tremendous discoveries
of gamma-ray bursts (GRBs) \citep[see][for a review]{mesz06,gehrels99}.
A surprising one is that large X-ray flares
are common in GRBs and occur at times well after the initial prompt
emission \citep{romano06,falcone07,chinc07,bernardini11}. The X-ray flare is
an episodic phenomena showing a sudden brightness at the late afterglow stage.
From the spectral and temporal analysis of X-ray flares, it is strongly
suggested that X-ray flares may have a common origin as the prompt gamma-ray
pulses and are related to the late time activity of the central
engine \citep{bernardini11,romano06}.

For the energy reservoir of
powering GRBs or X-ray flares, it is believed that they
are produced through an ultra-relativistic jet, i.e.,
a neutrino annihilation-driven jet or a Poynting flux-dominated jet.
For the former jet, neutrinos annihilate above
the disk and form a hot fireball which subsequently expand to accelerate
the jet by its thermal pressure \citep{pwf99,dimatteo02,chen07}.
On the other hand, in the case of a Poynting
flux-dominated jet, the jet derives its energy from the rotational energy
of the central black hole (BH) via the Blandford-Znajek (BZ) process
\citep{bz77,mckinney04,tchek08}.
Several mechanisms were proposed to explain the episodic phenomenon
of X-ray flares,
including fragmentation of a rapidly rotating core
\citep{king05}, magnetic regulation of the accretion flow \citep{pz06},
fragmentation of the accretion disk \citep{perna05}, differential
rotation in a post-merger millisecond pulsar \citep{dai06},
transition from thin to thick disk \citep{lazzati08},
He-synthesis-driven winds \citep{lee09}, the propagation instabilities
in GRB jets \citep{lazzati11}, and the episodic, magnetically dominated
jets \citep{yuan12}.
Apart from the energy and the episodic activity
of X-ray flares, their evolution has also been investigated, which shows that
the average energy released in the form of X-ray flares overlaid on
the power-law decay of the afterglow of GRBs
\citep{lazzati08,margutti11}.

In the present work, we will concentrate on another issue, i.e.,
whether or not the remnant disk after the prompt gamma-ray emission
can power X-ray flares through neutrino annihilation.
The luminosity of neutrino annihilation produced by the accretion disk
is sensitive to the accretion rate. In the late stage of disk evolution,
the accretion rate will probably be quite low
(e.g., $\sim 0.01 M_{\sun}$ s$^{-1}$) and the annihilation luminosity
will drop sharply according to previous numerical calculations
\citep[e.g.,][]{pwf99}. In other words, the observed X-ray flares,
if produced by the annihilation mechanism, will still require
a relatively high accretion rate. We take the second flare of GRB 070318
as an example, which has an isotropic luminosity of
$2.96 \times 10^{48}$ erg s$^{-1}$ and a rest frame duration
of 80.6 s (see Table~1). In order to produce such a luminosity by
neutrino annihilation, an accretion rate around $0.06 M_{\odot}$ s$^{-1}$
is required for a rapidly rotating BH ($a_{\ast} = 0.9$)
due to our numerical calculations in Section~3, which is consistent with
the results in \citet{pwf99}. Subsequently, the required remnant disk
mass is around $4.8 M_{\odot}$, which is obviously beyond what
the progenitor can provide \citep[e.g.,][]{shibata08,panna11}.

In this work, we will investigate the neutrino-cooled disk with the
magnetic coupling (MC) between the inner disk and the BH,
where the magnetic field connects the accretion disk and
the central BH \citep{li02,wang02,uzdensky05}. In such a model,
the angular momentum and the energy are transported from the BH horizon to
the accretion disk through a closed magnetic field. Accordingly,
this transported energy will be released and therefore increase
the radiation of the disk \citep{li02}. The MC effects
on the structure and radiation of neutrino-cooled disks have first
been studied by \citet{lei09}. They showed that both the neutrino
luminosity and the annihilation luminosity will increase significantly
owing to the MC process.
In the present work, we will focus on
relatively low accretion rates $0.001 \sim 0.1 M_{\sun}$ s$^{-1}$
to study the possibility of the remnant disk to power X-ray flares.
We would point out that, although such type of magnetic fields
is one of the possible field geometries discussed by
\citet[][Figure~2]{McK05} and has been studied by a few previous works,
the existence of such magnetic fields remains a problem according to
MHD simulations. We will discuss this issue in the last section.

The paper is organized as follows.
Equations are presented in Section~2.
The structure and neutrino radiation of the disk are calculated in Section~3.
A comparison of our numerical results with the observed
X-ray flares is shown in Section~4.
Conclusions and discussion are made in Section~5.

\section{Equations}

For neutrino-dominated accretion, the disk is extremely hot and dense
and the neutrino radiation can balance the
viscous dissipation. The structure and radiation of such a disk
have been widely investigated in previous works
\citep{pwf99,dimatteo02,gu06,chen07,liu07,lei09,pan12,liu13}.
Some simulations showed that the accretion flow is very dynamic and
the inner radius of the flow changes with time. In addition, the flow is
subject to various HD and MHD instabilities so that it is non-asymmetric.
In order to avoid the complexity of solving partial differential equations,
the present work is still based on the assumption of
a steady and axisymmetric accretion flow. In such case,
the basic equations of the neutrino-cooled accretion disk
including the MC effects may refer to
Section~2 of \citet{lei09}, and the relativistic effects
of the spinning BH were shown in \citet{riher95}.

For simplicity, we define the gravitational radius as
$r_{\rm g} \equiv GM/c^2$,
the dimensionless radius as $x \equiv r/r_{\rm g}$,
and the dimensionless spin parameter as $a_{\ast} \equiv cJ/GM^2$.
The disk is assumed to be Keplerian rotating, thus the angular velocity
of the flow is expressed as
\begin{equation}
\Omega = {c \over r} {1 \over (x^{1/2} + a_{\ast}x^{-1})} .
\label{eq:omegad}
\end{equation}

In the present work, we investigate the properties of
the neutrino-cooled disk with magnetic field lines connecting
the BH with the inner disk. Such a MC process
may have substantial effects on the energy and angular momentum balance
of the disk \citep{lipa00,li02,wang02,janiuk10,Kov11}.
In addition, \citet{uzdensky05} showed that the inner part of the
disk is magnetically coupled to the BH, but the magnetic field cannot
be stable in the outer region.
Here, we follow the assumptions of \citet{wang03} that the MC process
is constrained by a critical polar angle $\theta_0$, and the magnetic
field varies as a power law with the disk radius.
The magnetic torque exerted to the disk from the BH horizon
can be expressed as
\begin{equation}
T_{\rm MC} = 4 a_{\ast} \left( 1+\sqrt{1-a_{\ast}^2} \right) T_{0}
\int_{\theta_0}^{\pi/2} {(1-\beta)\sin^{3}\theta \over 2 -
(1 - \sqrt{1-a_{\ast}^{2}}) \sin^{2}\theta} \ d\theta , 
\label{eq:Tmc}
\end{equation}
where $\theta_0$ is a critical polar angle. In the scenario
\citep[e.g.,][Figure~1]{wang02}, the MC process exists
in the range $\theta_0 < \theta < \pi/2$. On the contrary, for the
space with $0 < \theta < \theta_0$, the BZ process may occur and
therefore some accreted materials may be pushed away, particularly
in the inner region of the disk. In the present study, we will
focus on the MC process and a constant accretion rate is assumed
for simplicity.
The critical angle is calculated by
\begin{equation}
\cos\theta_0 = \int_{1}^{\xi_{\rm out}}{\xi^{1-n}\chi^{2}_{\rm ms}
\sqrt{1+a_{\ast}^{2}\chi^{-4}_{\rm ms}\xi^{-2}
+ 2 a_{\ast}^{2}\chi^{-6}_{\rm ms}\xi^{-3}} \over
2 \sqrt{(1+a_{\ast}^{2}\chi^{-4}_{\rm ms}
+ 2 a_{\ast}^{2}\chi^{-6}_{\rm ms})(1 - 2 \chi^{-2}_{\rm ms}\xi^{-1}
+ a_{\ast}^{2}\chi^{-4}_{\rm ms}\xi^{-2})}} \ d\xi,
\end{equation}
where $\xi = r/r_{\rm ms}$, $\chi_{\rm ms} = \sqrt{r_{\rm ms}/r_{\rm g}}$,
$\xi_{\rm out} = r_{\rm out}/r_{\rm ms}$,
and $T_{0} = 3.26\times 10^{45} \left({B_{\rm H} / 10^{15}
{\rm G}}\right)^{2}
\left({M / M_{\odot}}\right)^{3}$ g cm$^2$ s$^{-2}$.
$\beta = \Omega /\Omega_{\rm H}$ is the ratio of the angular velocity
of the disk (Eq.~(\ref{eq:omegad}))
to the angular velocity at the horizon, where $\Omega_{\rm H}
= (ca_{\ast}/2r_{\rm g})/(1+\sqrt{1-a_{\ast}^2})$.
In addition, under the equipartition assumption
\citep[e.g.,][]{McK05,lei09},
the magnetic field strength at the horizon can be estimated as
$B_{\rm H}^2 = 8\pi c\dot M / r_{\rm g}^2$.

The energy equation is written as
\begin{equation}
Q^{+}_{\rm vis} = Q_G^+ + Q_{\rm MC}^+
= Q^{-}_{\rm adv} + Q^{-}_{\rm \nu} ,
\end{equation}
where $Q^{+}_{\rm vis}$ is the viscous heating rate, including
the contributions of the gravitational potential
$Q_G^+$ and the MC process
$Q_{\rm MC}^+ = - T_{\rm MC}/(4\pi r)\cdot d\Omega/dr$.
The quantities $Q^{-}_{\rm adv}$ and $Q^{-}_{\rm \nu}$ are respectively
the advective cooling rate and the neutrino cooling rate.
Here, we neglect the radiation of photons since they are trapped in the disk.
The neutrino cooling generally consists of the four processes:
the electron-positron pair annihilation,
the bremsstrahlung emission of nucleons,
the plasmon decay, and the Urca process \citep[e.g.,][]{liu07}.
In addition, for the equation of state,
the total pressure consists of five terms, i.e., the gas pressure,
the radiation pressure, the degeneracy pressure, the neutrino pressure,
and the magnetic pressure.
The detailed description of the neutrino cooling and the pressure can
be found in some previous papers \citep[e.g.,][]{pwf99,gu06,janiuk10}.
For the term of magnetic pressure, following \citet{lei09},
the ratio of magnetic pressure to total pressure is assumed to be 0.1.

\section{Numerical results}

\subsection{Disk structure}

In our calculations we fix $M = 3 M_{\odot}$, $\alpha = 0.1$,
$a_{\ast} = 0.9$, $n = 3$, and $r_{\rm out} = 100 r_{\rm g}$.
The numerical results of structure and radiation are shown in
Figure~1, where the solid lines and the dashed lines represent
the solutions with and without the MC process, respectively.
We choose three accretion rates, i.e.,
$\dot M = 0.005$, $0.05$, and $0.5$ $M_\odot$ s$^{-1}$ for the study.
Figures~1a, 1b, and 1c show respectively the radial profiles of
the mass density, the temperature, and the neutrino cooling efficiency
defined as the ratio of the neutrino cooling rate to
the viscous heating rate $Q_{\nu}^-/Q_{\rm vis}^+$.

It is seen that, for the outer region with $r \ga 20 r_{\rm g}$,
the solid and dashed lines are identical to each other, which
implies that the MC effects in this region are negligible.
On the contrary, for the inner region with $r \la 20 r_{\rm g}$,
the solid lines and the dashed lines are apparently separate, which
indicates that the MC effects
are substantial. For the inner region, Figure~1a shows that
the density with MC is obviously larger than that without MC.
The physical understanding is as follows. The MC process
can transfer angular momentum from the BH to the inner disk.
As a consequence, the additional transferred angular momentum will work as
a barrier to prevent the flow from radial acceleration, and therefore
the accreted matter will accumulate in this region and the mass density
will significantly increase. On the other hand,
for $\dot M = 0.5 M_{\sun}$ s$^{-1}$, it is shown by Figure~1a
that the solid line drops inwards even faster than the dashed line.
The reason is that, for large $\dot M$, the disk will become optically
thick to neutrinos. Thus, most of the generated neutrinos
are trapped in the disk instead of escaping away.
The total pressure, which includes the neutrino pressure, will therefore
increase significantly and the disk will probably
become geometrically thick. Accordingly, the mass density will drop sharply
due to the increased vertical height and radial velocity.
The neutrino trapping can be also indicated by Figure~1c, where the solid
line steeply drops inwards for $\dot M = 0.5 M_{\sun}$~s$^{-1}$,
indicating that most neutrinos are trapped in the disk
rather than being radiated.

\subsection{Annihilation luminosity}

In the scenario of hyper-accretion disks, the GRBs are powered by the
neutrino and anti-neutrino annihilation above the surface of disks.
Since the X-ray flares and the gamma-ray emission have the same
power origin,
the flares may still be powered by the neutrino annihilation.
The disk luminosity is calculated from the marginal stable orbit
$r_{\rm ms}$ to the outer boundary $r_{\rm out}$.
The neutrino luminosity from the disk is calculated as
\begin{equation}
\label{eq:lnu}
L_\nu = 4\pi \int_{r_{\rm ms} }^{r_{\rm out}} r Q_{\nu}^- \ dr . 
\end{equation}
The total annihilation luminosity is calculated by the integration
over the whole space outside the disk, following the method in
previous works \citep[e.g.,][]{ruff97,pwf99}.

Figure 2 shows the variations of the neutrino luminosity $L_{\nu}$
($L_{\nu}^{'}$) and the annihilation luminosity $L_{\nu \bar{\nu}}$
($L_{\nu \bar{\nu}}^{'}$) with the mass accretion rate $\dot M$.
Similar to Figure~1, the solid lines and the dashed lines represent
the solutions with and without the MC process, respectively.
The upper solid (dashed) line corresponds to $L_{\nu}$ ($L_{\nu}^{'}$),
and the lower solid (dashed) line corresponds to $L_{\nu \bar{\nu}}$
($L_{\nu \bar{\nu}}^{'}$). It is seen that, both the neutrino luminosity
and the annihilation luminosity with the MC process are significantly
larger than those without the MC process. In particular for
relatively low accretion rates such as $\dot M \sim 0.01 M_{\sun}$ s$^{-1}$,
$L_{\nu \bar{\nu}}$ is larger than $L_{\nu \bar{\nu}}^{'}$ by up to
four orders of magnitude. 
The physical reason is that, in addition to the gravitational energy,
the MC process can efficiently extract
the BH rotational energy into neutrino radiation.
Moreover, the apparent difference between the lower solid line and
the lower dashed line implies that it is quite possible for
a remnant low-mass disk to power X-ray flares.

\subsection{Efficiency of BH rotational energy extraction}

We compare the efficiency of BH rotational energy extraction between
the MC process and the BZ mechanism.
The former efficiency is defined as
\begin{equation}
\label{eq:etaMC}
\eta_{\rm MC} = \frac{\int_{r_{\rm ms}}^{r_{\rm out}}
4\pi r Q_{\rm MC}^{+} \ dr}{\dot M c^2} \ .
\end{equation}
For the BZ mechanism, the magnetic field lines, which are dragged in by the
accretion disk, accumulate around the BH horizon and then get twisted
by the BH space-time, which enable the extraction of BH rotational energy.
The efficiency of the BZ process was discussed in \citet{tchek11},
\begin{equation}
\label{eq:eta}
\eta_{\rm BZ} = \frac{\kappa}{4\pi c} \left(\frac{\Omega_{\rm H} r_g}{c}
\right)^2 \Phi_{0}^2 f(\Omega_{\rm H}) \ ,
\end{equation}
where $\Phi_{0} = \Phi_{\rm BH}/(\dot{M} c r_g^2)^{1/2}$
is the dimensionless magnetic flux threading the BH,
$\kappa$ is a numerical constant related to
the magnetic field geometry (we adopt $\kappa = 0.044$ here),
and $f(\Omega_{\rm H})$ is set to be 1.
The above equation shows that the BZ efficiency is
relevant to the dimensionless magnetic flux $\Phi_{0}$.
The value of $\Phi_{0}$, however, is quite uncertain and may be related to
the accretion rate. It can be regarded as the ability for
the accretion disk to drag magnetic fields into the BH horizon.
Recent simulations show that $\Phi_{0}$ can be as large as several
tens and the disk can be depicted as a magnetic arrest disk.
At a high $\Phi_{0}$, a large amount of magnetic flux is transported
to the center and the efficiency can be larger than 1. Here
we simply assume a constant magnetic flux, $\Phi_{0} = 50$, which is
close to the simulation result $\Phi_{0} \approx 47$ \citep{tchek11}. 

Equations~(\ref{eq:etaMC}) implies that $\eta_{\rm MC}$ is independent
of the accretion rate $\dot M$.
The variation of $\eta_{\rm MC}$ and $\eta_{\rm BZ}$ with the spin
parameter $a_{\ast}$ is shown in Figure~3.
It is seen that,
for $a_{\ast} \ga 0.5$, $\eta_{\rm MC}$ can be significantly larger
than the efficiency of normal accretion process $0.06 \la \eta \la 0.42$,
and can be even larger than 1 for extremely spinning BH.
Such a result indicates that the MC
process is quite an efficient mechanism to
extract the BH rotational energy. The efficiency
$\eta_{\rm MC}$ decreases sharply with decreasing $a_{\ast}$ for
$a_{\ast} \la 0.4$. The reason is that the angular
velocity of BH will be less than that of the disk at ISCO
for $a_{\ast} \la 0.36$, and therefore
the MC process will become weak for BH spin below this
critical value. 
It is also seen that, for $a_{\ast} \ga 0.4$, $\eta_{\rm MC}$ is
several times larger than $\eta_{\rm BZ}$, whereas for $a_{\ast} \la 0.3$,
$\eta_{\rm BZ}$ is significantly larger than $\eta_{\rm MC}$ due to
a sharp decrease of $\eta_{\rm MC}$ with decreasing $a_{\ast}$.
In other words, for fast spinning BH systems, MC may be more powerful
than BZ on the extraction of rotational energy, whereas for
slow spinning BH systems, BZ is likely to be more powerful.

Another different effect between MC and BZ is that,
the BZ process can directly transfer the rotational energy
into the jet, whereas the MC process can only transfer the rotational
energy into the disk, and the jet power is also related to another
process, i.e., neutrino annihilation. Thus, for the efficiency of
powering the jet, the BZ process will probably be much more efficient.
For a comparison, the dotted line shows the efficiency of radiation
due to neutrino annihilation $\eta_{\nu\bar\nu}$
($\equiv L_{\nu\bar\nu}/\dot{M}c^2$) for a typical accretion
rate $\dot M = 0.05$~$M_{\sun}$~s$^{-1}$. It is seen that
$\eta_{\rm BZ}$ is significantly larger than $\eta_{\nu\bar\nu}$
for any $a_{\ast}$.
Thus, the BZ process is more efficient to power a jet than the MC process.
It is also possible for the BZ process to work as the central engine
to power the X-ray flares.

\section{Comparison with observations}

In order to compare our numerical results with the observations
of X-ray flares, we compile a sample of 21 GRBs with 43 flares as shown
in Table~1, which includes all the flares with available redshift
in Table~1 of \citet{chinc10}.
The width of the flares is calculated in the rest frame:
$w_{\rm res} = w/(1+z)$, where $z$ is the redshift,
$w$ is the observed width,
and $w_{\rm res}$ is the width in the rest frame.
The isotropic energy of a flare $E_{\rm flare}$ can be estimated
from the fluence $S$: $E_{\rm flare} = 4\pi D_l^2 S/(1+z)$,
where $D_l$ is the luminosity distance.
Thus, the average, isotropic luminosity can be
obtained as $L_{\rm iso} = E_{\rm flare}/w_{\rm res}$.
 
A comparison of our numerical results with the observations
is shown in Figure~\ref{fig:flare}, which includes all the
flares in Table~1. Similar to Figures~1 and 2, the solid lines
represent the results with MC, whereas the dashed
lines represent the results without MC.
The upper and lower solid (dashed) lines correspond to the disk mass
$M_{\rm disk} = 0.5$~$M_{\sun}$ and 0.05~$M_{\sun}$, respectively.
The four theoretical lines are calculated by
$w_{\rm res} = M_{\rm disk}/\dot M$ together with the relationship
between $\dot M$ and $L_{\nu \bar\nu}$ obtained in Section~3.

Figure~\ref{fig:flare} shows that most of the flares locate
above the upper dashed line,
which means that for the case without MC,
it will generally require a remnant disk mass $M_{\rm disk}$ significantly
larger than 0.5~$M_{\sun}$, which may be unpractical.
However, it is seen that most of the flares exist
between the two solid lines, which indicates that
if MC works, a reasonable remnant disk
with $0.05 M_{\sun} \la M_{\rm disk} \la 0.5 M_{\sun}$ is able to power
nearly all the plotted flares. As mentioned in Section~1, taking
the second flare of GRB 070318 as an example,
the model without MC
requires $M_{\rm disk} \approx 4.8 M_{\sun}$. On the contrary,
if the MC effects are taken into account, the
luminosity of $2.96 \times 10^{48}$ erg s$^{-1}$ corresponds to
$\dot M \approx 0.0032 M_{\sun}$ s$^{-1}$ (according to our
numerical calculations shown by the lower solid line in Figure~2).
Therefore, the rest frame duration 80.6 s only requires
$M_{\rm disk} \approx 0.26 M_{\sun}$ for powering this flare.

\section{Conclusions and discussion}

In this paper, we have studied the neutrino-cooled disks by taking
into account the MC process between the central BH and the inner disk.
We have shown that, for mass accretion rates around
$0.001 \sim 0.1$ $M_{\sun}$ s$^{-1}$, the luminosity of neutrino annihilation
can be enhanced by up to four orders of magnitude due to
the MC effects. As a consequence, the remnant disk with
$M_{\rm disk} \la 0.5$~$M_{\sun}$ may power most of the observed X-ray
flares with the rest frame duration less than 100 seconds.

We would point out that, for a few X-ray flares with extremely 
long duration, the neutrino-cooled disk cannot work as the central
engine, even though MC is included.
For example, GRB 050724 has three flares with a redshift of 0.258.
The third flare of this source has a rest frame width of $3.1\times 10^5$
seconds and the luminosity $6.7 \times 10^{43}$ erg s$^{-1}$.
According to our numerical results in Figure 2,
the accretion rate should be around $1.7 \times 10^{-4} M_{\odot}$ s$^{-1}$
for the case including MC. Thus, it requires
a remnant disk $M_{\rm disk} > 50 M_{\odot}$, which is obviously unphysical.
On the other hand, X-ray flares with peak time less than and larger than
1000 s may have different origin \citep{margutti11}.
The mechanism for powering the flares with peak time larger than
1000 s is worthy for further studies, but is beyond the scope of
the present paper.

Another issue we would like to stress is related to the configuration
of the large-scale magnetic field in accretion disks.
Although the MC process has been studied by quite a few previous works,
such type of magnetic fields, however, has not been found in
MHD simulations. Thus, it remains unclear whether the MC process
can occur between the inner disk and the central BH.
On the other hand, some simulations showed that
the BZ mechanism can be a solution to the GRB's central engine
\citep[e.g.,][]{tchek08}.
In the scenario of the Poynting flux-dominated jet,
the efficiency of extracting the BH rotational energy
mainly depends on the magnetic flux being dragged in
\citep{tchek11,mckinney12}.
The theoretical analysis showed that it requires a geometrically
thick disk to transport a large mount flux into the center
\citep{lubow94,rothstein08,beckwith09,cao11}.
Simulations also confirmed that a thick disk can
efficiently transport magnetic flux \citep{mckinney12}.
For a neutrino-cooled disk, neutrinos play a vital role
to release the dissipation heat and the disk is likely
to be geometrically thin \citep[e.g.,][]{shibata07}.
Then, for the Poynting flux-dominated jet, it remains a problem
whether the accretion flow can accumulate adequate
magnetic fields to the inner region to power the jet.

In this work, the flow is assumed to be steady and
the mass accretion rate is a free parameter. In other words,
for a given accretion rate, we will obtain a corresponding solution.
On the other hand,
the simulations of \citet{tchek11} found a correlated variation
between the accretion rate and the magnetic flux $\Phi_{\rm BH}$.
In this spirit, a varying strength of MC process may also have effects on
the variation of accretion rate. Such a study requires further
time-dependent calculations.

\acknowledgments

We thank Raffaella Margutti, Shujin Hou, Da-Bin Lin, and Mou-Yuan Sun
for beneficial discussions, and the referee for helpful suggestions.
This work was supported by the National Natural Science Foundation
of China under grants 11073015, 11103015, 11222328, and 11233006.

\clearpage

\begin{figure}
\centering
\includegraphics[width=10cm]{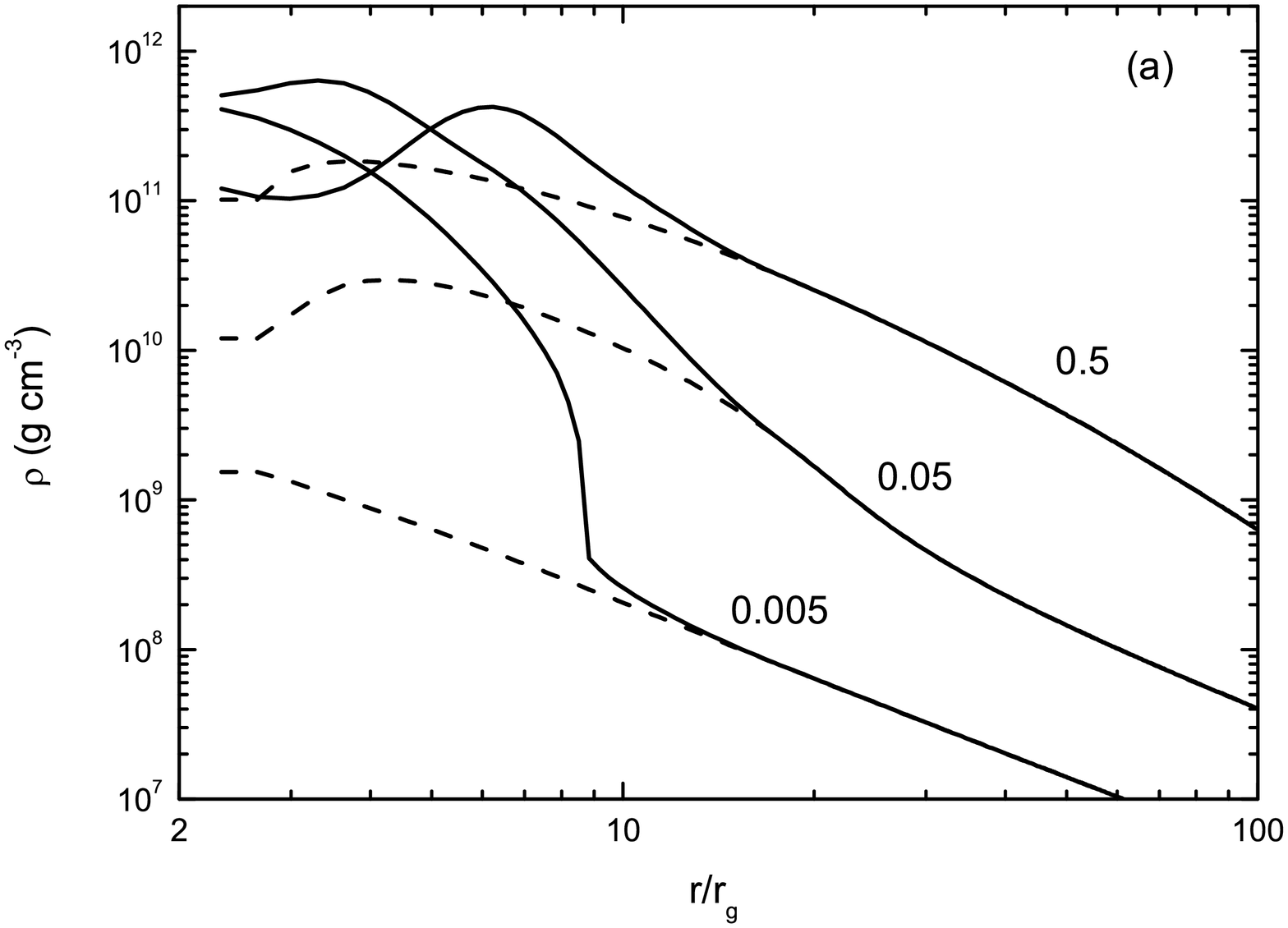}
\includegraphics[width=10cm]{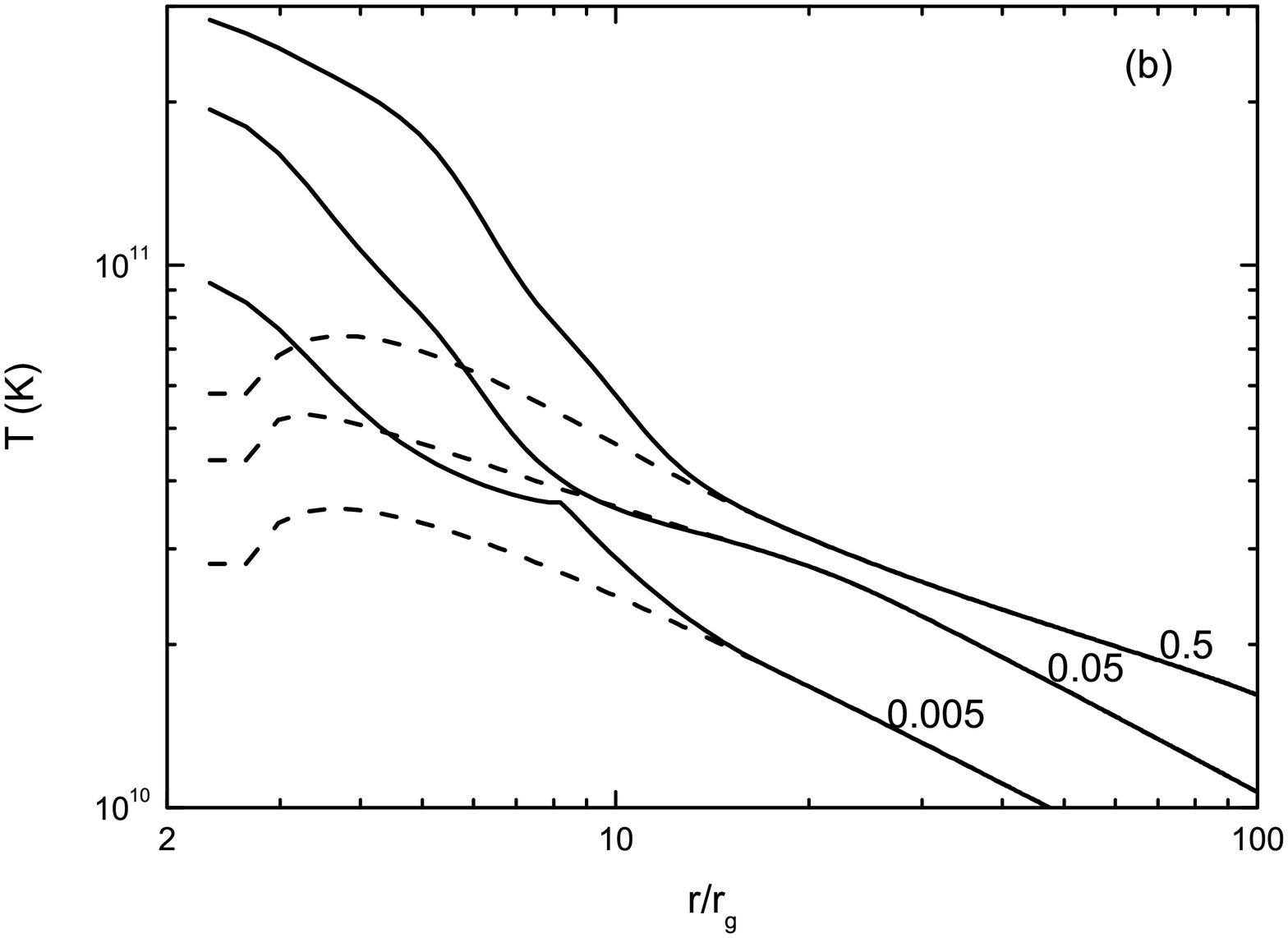}
\includegraphics[width=10cm]{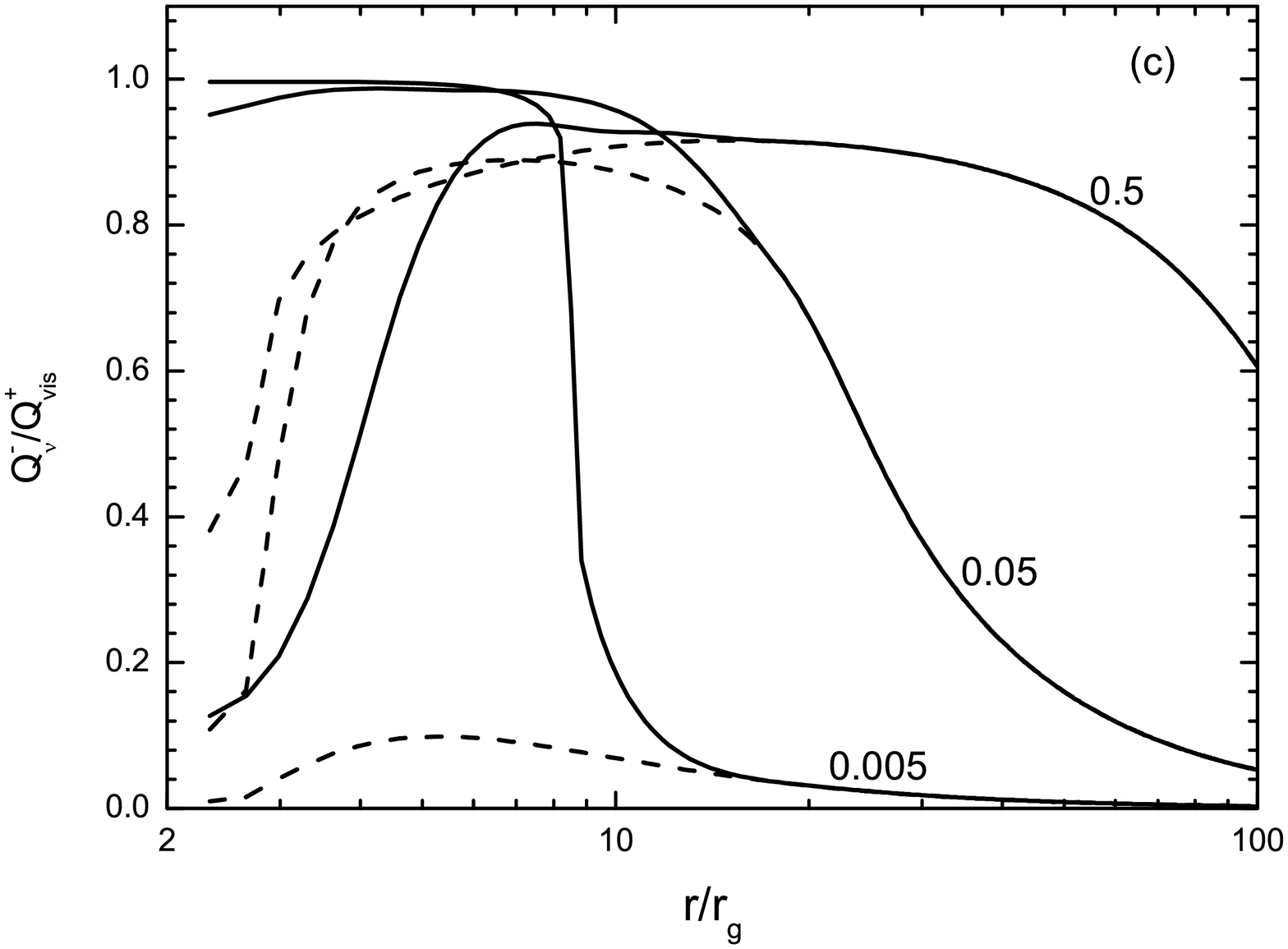}
\caption{
Radial profiles of the density $\rho$, the temperature $T$, and the ratio
of the neutrino cooling to the viscous heating $Q_{\nu}^{-}/Q_{\rm vis}^{+}$
for $\dot M = 0.005$, 0.05, 0.5 $M_\odot$ s$^{-1}$.
The solid lines and the dashed lines represent the solutions
with and without the MC process, respectively.
}
\label{fig1}
\end{figure}

\clearpage

\begin{figure}
\plotone{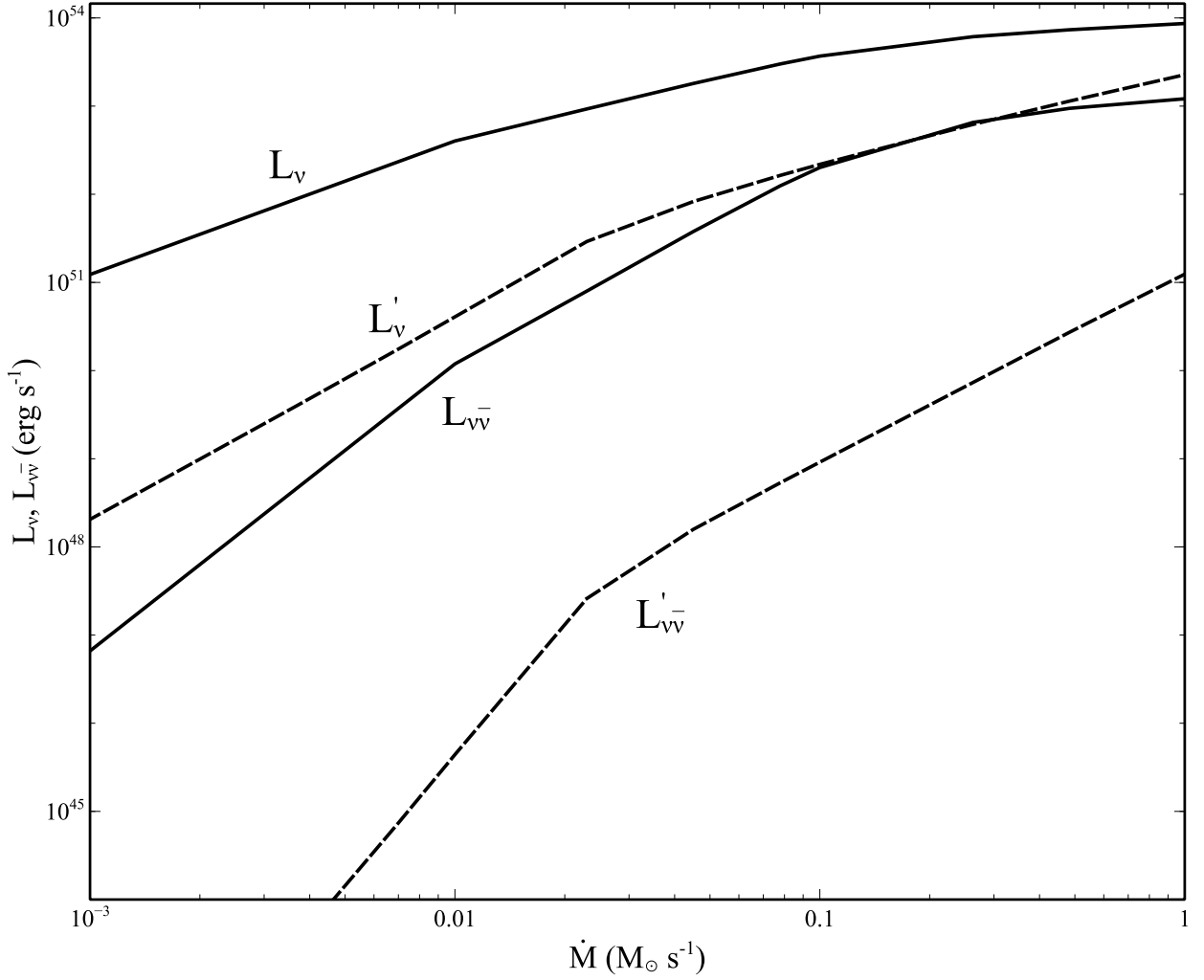}
\caption{
Variations of the neutrino luminosity $L_{\nu}$ ($L_{\nu}^{'}$)
and the annihilation luminosity $L_{\nu \bar{\nu}}$ ($L_{\nu \bar{\nu}}^{'}$)
with the mass accretion rate.
The upper and lower solid lines correspond to $L_{\nu}$
and $L_{\nu \bar{\nu}}$ with the MC process, respectively,
whereas the upper and lower dashed lines correspond to $L_{\nu}^{'}$ and
$L_{\nu \bar{\nu}}^{'}$ without the MC process, respectively.
}
\label{fig:lann}
\end{figure}

\clearpage

\begin{figure}
\plotone{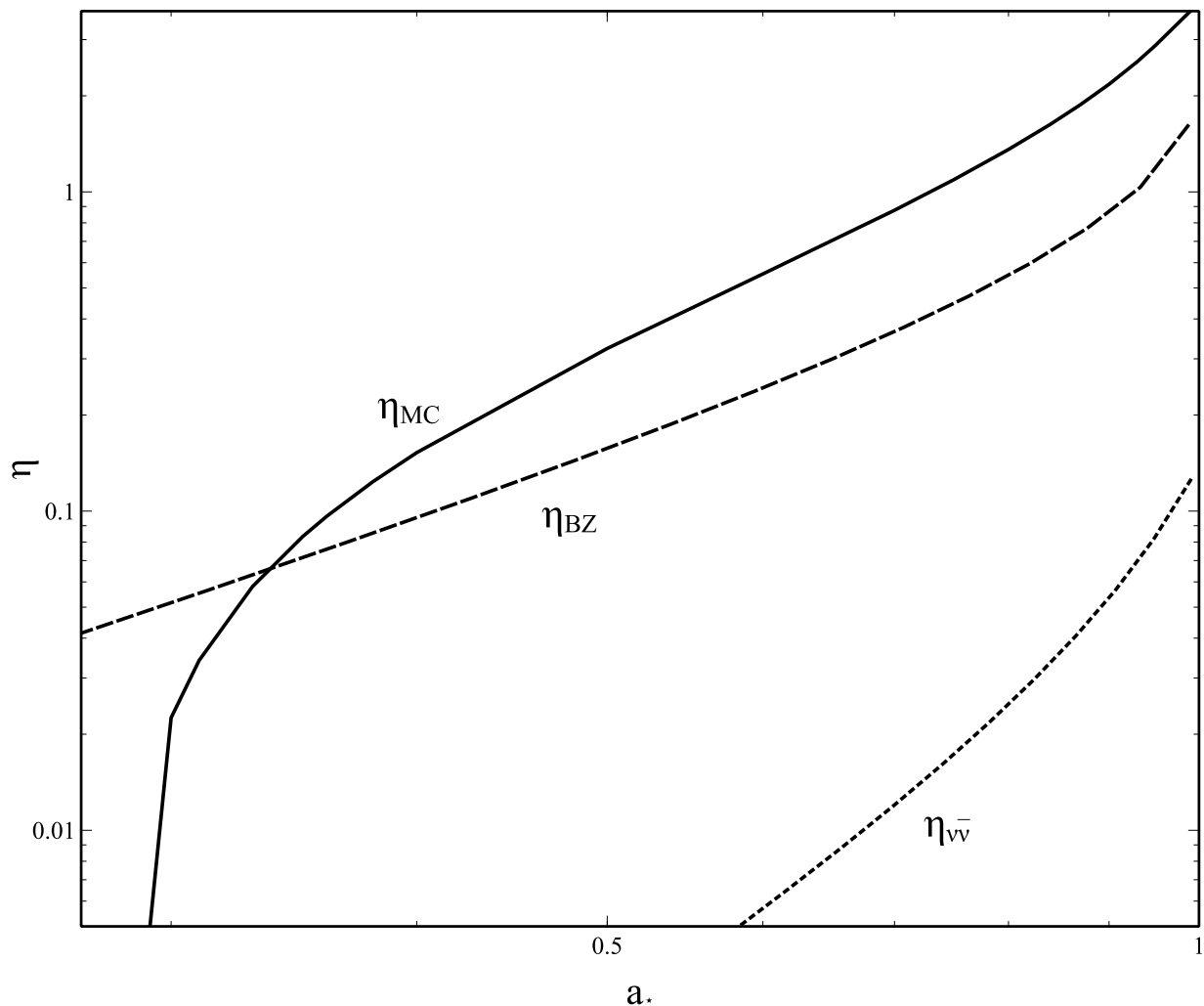}
\caption{
Variation of the efficiency $\eta$ with the spin parameter $a_{\ast}$.
The solid line shows the efficiency of the energy transfer from
the rotating BH to the disk by the MC process.
For a comparison, the dashed line shows the efficiency of the
energy extraction by the BZ mechanism with $\Phi_{0} = 50$.
The dotted line shows the efficiency of radiation
due to neutrino annihilation for a typical accretion rate
$\dot M = 0.05$~$M_{\sun}$~s$^{-1}$.
}
\label{fig:effi}
\end{figure}

\clearpage

\begin{figure}
\plotone{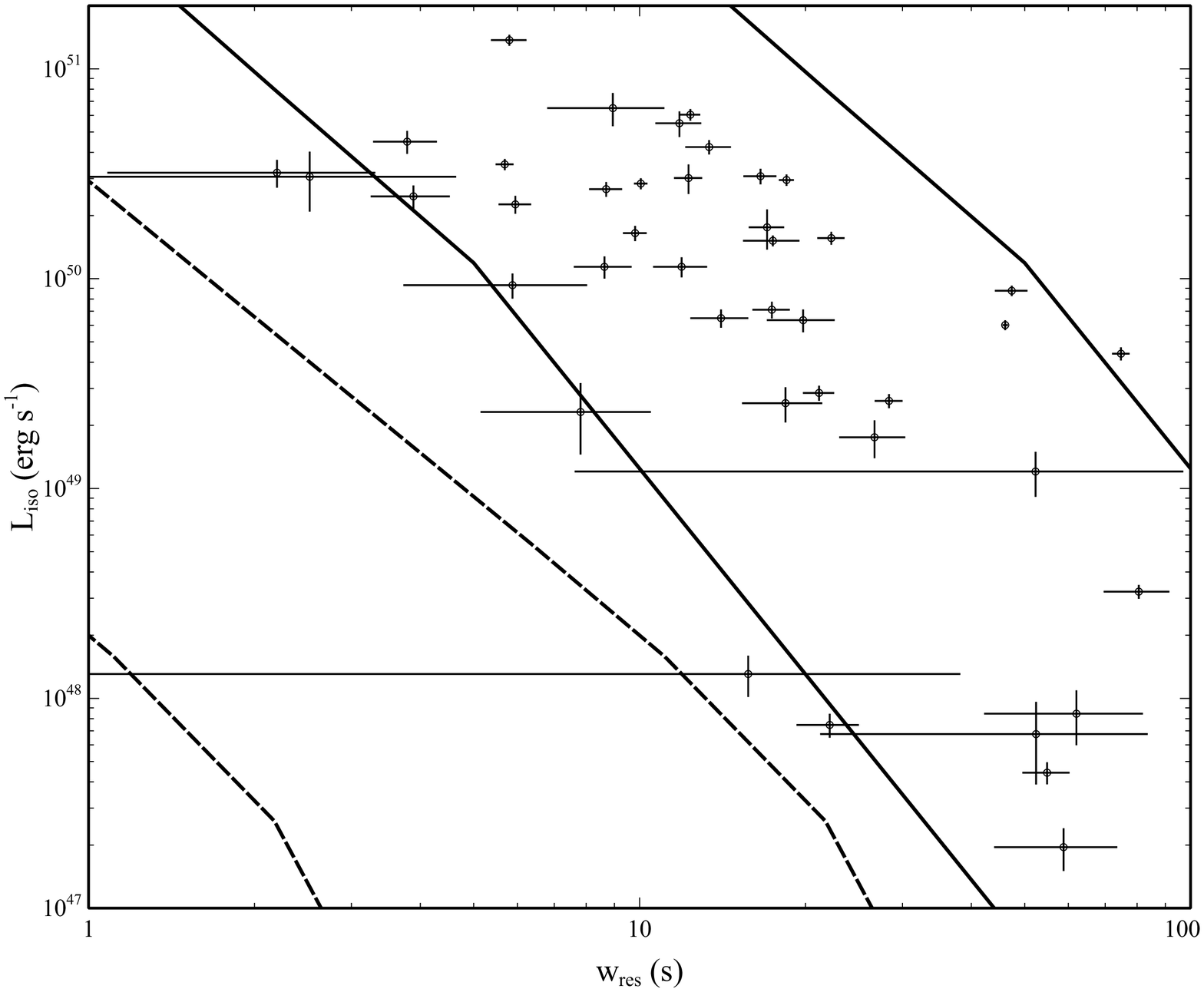}
\caption{
A comparison of our numerical results with the observations in the
$L_{\rm iso}-w_{\rm res}$ diagram.
The upper and lower solid lines correspond to the remnant disk mass
$M_{\rm disk} = 0.5 M_{\odot}$ and $0.05 M_{\odot}$
with the MC process, respectively,
whereas the upper and lower dashed lines correspond to
$M_{\rm disk} = 0.5 M_{\odot}$ and $0.05 M_{\odot}$ without
the MC process, respectively.
The data for the 43 flares are shown in Table~1.
}
\label{fig:flare}
\end{figure}

\clearpage

\begin{deluxetable}{cccc}
\tabletypesize{\scriptsize}
\tablecaption{Observed flares with available redshift}
\tablewidth{0pt}
\tablehead{
\colhead{GRB} & \colhead{z} & \colhead{$L_{\rm iso}$($10^{50}$erg s$^{-1}$)}
& \colhead{$w_{\rm res}$(s)}
}
\startdata
050730  &  3.967  &  1.0210 $\pm$ 0.4929  &  8.6370 $\pm$ 1.0268  \\
050730  &  3.967  &  1.3644 $\pm$ 0.3684  &  22.2871 $\pm$ 1.2684  \\
050730  &  3.967  &  0.5679 $\pm$ 0.2609  &  19.8108 $\pm$ 2.7783  \\
050908  &  3.344  &  0.1648 $\pm$ 0.0520  &  26.6805 $\pm$ 3.6602  \\
060115  &  3.53  &  0.2332 $\pm$ 0.0960  &  18.4106 $\pm$ 3.0464  \\
060210  &  3.91  &  5.5296 $\pm$ 0.9617  &  12.3625 $\pm$ 0.5295  \\
060210  &  3.91  &  2.9090 $\pm$ 0.5818  &  12.2607 $\pm$ 0.7128  \\
060418  &  1.489  &  2.6893 $\pm$ 0.2241  &  10.0442 $\pm$ 0.2812  \\
060512  &  0.4428  &  0.0041 $\pm$ 0.0010  &  54.8933 $\pm$ 5.4062  \\
060526  &  3.221  &  12.0001 $\pm$ 14.6252  &  5.8043 $\pm$ 0.4264  \\
060526  &  3.221  &  6.0765 $\pm$ 2.6736  &  8.9552 $\pm$ 2.1559  \\
060526  &  3.221  &  4.9712 $\pm$ 1.1600  &  11.8218 $\pm$ 1.1372  \\
060526  &  3.221  &  1.6612 $\pm$ 0.4345  &  17.0339 $\pm$ 1.2556  \\
060604  &  2.68  &  2.3969 $\pm$ 0.6453  &  8.6957 $\pm$ 0.5978  \\
060604  &  2.68  &  2.1283 $\pm$ 0.3502  &  5.9511 $\pm$ 0.4076  \\
060607A  &  3.082  &  2.2870 $\pm$ 0.7277  &  3.8951 $\pm$ 0.6369  \\
060607A  &  3.082  &  2.6306 $\pm$ 0.3617  &  18.4713 $\pm$ 0.5879  \\
060707  &  3.425  &  0.2153 $\pm$ 0.1846  &  7.8192 $\pm$ 2.6667  \\
060714  &  2.711  &  4.0362 $\pm$ 0.5259  &  13.3657 $\pm$ 1.2665  \\
060714  &  2.711  &  2.8113 $\pm$ 0.9618  &  2.2096 $\pm$ 1.1048  \\
060714  &  2.711  &  4.1165 $\pm$ 1.0400  &  3.7726 $\pm$ 0.5120  \\
060714  &  2.711  &  3.1626 $\pm$ 0.5750  &  5.6858 $\pm$ 0.2156  \\
060729  &  0.54  &  0.2463  $\pm$ 0.0265  &  28.3766 $\pm$ 1.6234  \\
060814  &  0.84  &  0.2704  $\pm$ 0.0419  &  21.1957 $\pm$ 1.3587  \\
060904B  &  0.703  &  0.5598 $\pm$ 0.0336  &  46.0951 $\pm$ 0.5285  \\
060908  &  1.8836  &  0.0066 $\pm$ 0.0066  &  52.4345 $\pm$ 31.1416  \\
060908  &  1.8836  &  0.0070 $\pm$ 0.0098  &  62.0752 $\pm$ 19.8710  \\
070306  &  1.4959  &  0.6855 $\pm$ 0.1143  &  17.3885 $\pm$ 1.3622  \\
070318  &  0.836  &  0.0116 $\pm$ 0.0128  &  15.7407 $\pm$ 22.4401  \\
070318  &  0.836  &  0.0296 $\pm$ 0.0043  &  80.5556 $\pm$ 11.0022  \\
070721B  &  3.626  &  0.6013 $\pm$ 0.8718  &  26.3078 $\pm$ 63.3593  \\
070721B  &  3.626  &  2.7102 $\pm$ 1.8763  &  2.5292 $\pm$ 2.1185  \\
070721B  &  3.626  &  0.8519 $\pm$ 0.3139  &  5.8798 $\pm$ 2.1617  \\
070721B  &  3.626  &  0.1109 $\pm$ 0.0907  &  52.3130 $\pm$ 44.6822  \\
070724A  &  0.457  &  0.0067 $\pm$ 0.0058  &  22.1688 $\pm$ 2.8826  \\
070724A  &  0.457  &  0.0018 $\pm$ 0.0009  &  58.8195 $\pm$ 14.8250  \\
071031  &  2.692  &  2.7287 $\pm$ 1.0720  &  16.5764 $\pm$ 1.1376  \\
071031  &  2.692  &  1.0844 $\pm$ 0.2575  &  11.9177 $\pm$ 1.3272  \\
071031  &  2.692  &  0.6206 $\pm$ 0.1494  &  14.0574 $\pm$ 1.7064  \\
071031  &  2.692  &  0.4104 $\pm$ 0.0410  &  74.7833 $\pm$ 2.7086  \\
080210  &  2.641  &  1.4834 $\pm$ 0.3350  &  9.8050 $\pm$ 0.4944  \\
080310  &  2.42  &  0.8247 $\pm$ 0.0654  &  47.3684 $\pm$ 3.2164  \\
080310  &  2.42  &  1.3890 $\pm$ 0.1543  &  17.4561 $\pm$ 2.0468  \\
\enddata
\end{deluxetable}


\begin{thebibliography}{}

\bibitem[Beckwith et al.(2009)]{beckwith09}
Beckwith, K., Hawley, J.~F., \& Krolik, J.~H.\ 2009, \apj, 707, 428

\bibitem[Bernardini et al.(2011)]{bernardini11}
Bernardini, M.~G., Margutti, R., Chincarini, G.,
Guidorzi, C., \& Mao, J.\ 2011, \aap, 526, A27 

\bibitem[Blandford \& Znajek(1977)]{bz77}
Blandford, R.~D., \& Znajek, R.~L.\ 1977, \mnras, 179, 433

\bibitem[Cao(2011)]{cao11}
Cao, X.\ 2011, \apj, 737, 94

\bibitem[Chen \& Beloborodov(2007)]{chen07}
Chen, W.-X., \& Beloborodov, A.~M.\ 2007, \apj, 657, 383 

\bibitem[Chincarini et al.(2007)]{chinc07}
Chincarini, G., Moretti, A., Romano, P., et al.\ 2007, \apj, 671, 1903 

\bibitem[Chincarini et al.(2010)]{chinc10}
Chincarini, G., Mao, J., Margutti, R., et al.\ 2010, \mnras, 406, 2113 

\bibitem[Dai et al.(2006)]{dai06}
Dai, Z.~G., Wang, X.~Y., Wu, X.~F., \& Zhang, B.\ 2006, Science, 311, 1127 

\bibitem[Di Matteo et al.(2002)]{dimatteo02}
Di Matteo, T., Perna, R., \& Narayan, R.\ 2002, \apj, 579, 706 

\bibitem[Falcone et al.(2007)]{falcone07}
Falcone, A.~D., Morris, D., Racusin, J., et al.\ 2007, \apj, 671, 1921 

\bibitem[Gehrels et al.(2009)]{gehrels99}
Gehrels, N., Ramirez-Ruiz, E., \& Fox, D.~B.\ 2009, \araa, 47, 567 

\bibitem[Gu et al.(2006)]{gu06}
Gu, W.-M., Liu, T., \& Lu, J.-F.\ 2006, \apjl, 643, L87 

\bibitem[Janiuk \& Yuan(2010)]{janiuk10}
Janiuk, A., \& Yuan, Y.-F.\ 2010, \aap, 509, A55 

\bibitem[King et al.(2005)]{king05}
King, A. R., et al. 2005, \apj, 630, L113 

\bibitem[Kov\'{a}cs et al.(2011)]{Kov11}
Kov\'{a}cs, Z., Gergely, L., \& Biermann, P. L. 2011, \mnras, 416, 991

\bibitem[Lazzati et al.(2011)]{lazzati11}
Lazzati, D., Blackwell, C.~H., Morsony, B.~J., \& Begelman, M.~C.
2011, \mnras, 411, L16

\bibitem[Lazzati et al.(2008)]{lazzati08}
Lazzati, D., Perna, R., \& Begelman, M. C. 2008, \mnras, 388, L15

\bibitem[Lee et al.(2009)]{lee09}
Lee, W. H., Ramirez-Ruiz, E., \& L\'{o}pez-C\'{a}mara D. 2009, \apj, 699, L93

\bibitem[Lei et al.(2009)]{lei09}
Lei, W. H., Wang, D. X., Zhang, L., et al.\ 2009, \apj, 700, 1970 

\bibitem[Li(2002)]{li02}
Li, L.-X.\ 2002, \apj, 567, 463 

\bibitem[Li \& Paczy{\'n}ski(2000)]{lipa00}
Li, L.-X., \& Paczy{\'n}ski, B.\ 2000, \apjl, 534, L197 

\bibitem[Liu et al.(2007)]{liu07}
Liu, T., Gu, W.-M., Xue, L., \& Lu, J.-F.\ 2007, \apj, 661, 1025

\bibitem[Liu et al.(2013)]{liu13}
Liu, T., Xue, L., Gu, W.-M., \& Lu, J.-F.\ 2013, \apj, 762, 102

\bibitem[Lubow et al.(1994)]{lubow94}
Lubow, S.~H., Papaloizou, J.~C.~B., \& Pringle, J.~E.\ 1994,
\mnras, 267, 235

\bibitem[Margutti et al.(2011)]{margutti11}
Margutti, R., Bernardini, G., Barniol Duran, R., et al. 2011, \mnras, 410, 1064 

\bibitem[McKinney(2005)]{McK05}
McKinney, J. C. 2005, \apj, 630, L5

\bibitem[McKinney \& Gammie(2004)]{mckinney04}
McKinney, J.~C., \& Gammie, C.~F.\ 2004, \apj, 611, 977

\bibitem[McKinney et al.(2012)]{mckinney12}
McKinney, J.~C., Tchekhovskoy, A., \& Blandford, R.~D.\ 2012,
\mnras, 423, 3083

\bibitem[M{\'e}sz{\'a}ros(2006)]{mesz06}
M{\'e}sz{\'a}ros, P.\ 2006, Reports on Progress in Physics, 69, 2259 

\bibitem[Pan \& Yuan(2012)]{pan12}
Pan, Z., \& Yuan, Y.-F. 2012, \apj, 759, 82

\bibitem[Pannarale et al.(2011)]{panna11}
Pannarale, F., Tonita, A., \& Rezzolla, L.\ 2011, \apj, 727, 95

\bibitem[Perna et al.(2006)]{perna05}
Perna, R., Armitage, P.~J., \& Zhang, B.\ 2006, \apjl, 636, L29 

\bibitem[Popham et al.(1999)]{pwf99}
Popham, R., Woosley, S.~E., \& Fryer, C.\ 1999, \apj, 518, 356

\bibitem[Proga \& Zhang(2006)]{pz06}
Proga, D., \& Zhang, B.\ 2006, \mnras, 370, L61 

\bibitem[Riffert \& Herold(1995)]{riher95}
Riffert, H., \& Herold, H.\ 1995, \apj, 450, 508

\bibitem[Romano et al.(2006)]{romano06}
Romano, P., Moretti, A., Banat, P.~L., et al.\ 2006, \aap, 450, 59 

\bibitem[Rothstein \& Lovelace(2008)]{rothstein08}
Rothstein, D.~M., \& Lovelace, R.~V.~E.\ 2008, \apj, 677, 1221

\bibitem[Ruffert et al.(1997)]{ruff97}
Ruffert, M., Janka, H. -T., Takahashi, K., \& Schaefer, G.
1997, \aap, 319, 122

\bibitem[Shibata et al.(2007)]{shibata07}
Shibata, M., Sekiguchi, Y.-I., \& Takahashi, R.\ 2007,
Progress of Theoretical Physics, 118, 257

\bibitem[Shibata \& Taniguchi(2008)]{shibata08}
Shibata, M., \& Taniguchi, K.\ 2008, \prd, 77, 084015

\bibitem[Tchekhovskoy et al.(2008)]{tchek08}
Tchekhovskoy, A., McKinney, J.~C., \& Narayan, R.\ 2008, \mnras, 388, 551

\bibitem[Tchekhovskoy et al.(2011)]{tchek11}
Tchekhovskoy, A., Narayan, R., \& McKinney, J.~C.\ 2011, \mnras, 418, L79

\bibitem[Uzdensky(2005)]{uzdensky05}
Uzdensky, D.~A.\ 2005, \apj, 620, 889 

\bibitem[Wang et al.(2002)]{wang02}
Wang, D. X., Xiao, K., \& Lei, W. H.\ 2002, \mnras, 335, 655 

\bibitem[Wang et al.(2003)]{wang03}
Wang, D. X., Lei, W. H., \& Ma, R. Y.\ 2003, \mnras, 342, 851 

\bibitem[Yuan \& Zhang(2012)]{yuan12}
Yuan, F., \& Zhang, B. 2012, \apj, 757, 56

\end{thebibliography}
\end{document}